\documentstyle[12pt,epsf,axodraw]{article}

\large
\newcommand{\be}{\begin{eqnarray}}
\newcommand{\ee}{\end{eqnarray}}
\newcommand\del{\partial}
\newcommand\dels{\partial \hspace{-0.23cm} /}
\newcommand\pslash{p \hspace{-0.20cm} /}
\newcommand\Dels{D \hspace{-0.26cm} /}

\newcommand\noi {\noindent}

\begin{document}
\setlength{\baselineskip}{21pt}
\pagestyle{empty}
\vfill
\eject
\begin{flushright} USITP-99-01 \\
February 1999
\end{flushright}
\mbox{}
\vskip 1.5cm
\centerline{\Large\bf Thermal Effects on the Low Energy}
\centerline{\Large\bf $N=2$ SUSY Yang--Mills Theory}
\vskip 1.2 cm
\begin{center}
{\bf J. Wirstam}  \\
Institute of Theoretical Physics\\
University of Stockholm \\
Box 6730, S-113 85 Stockholm, Sweden \\
\end{center}

\vskip 1cm\noi
\centerline{\bf ABSTRACT}
\vskip .5cm
Using the low energy effective action of the $N=2$ supersymmetric
SU(2) Yang--Mills theory  we calculate the free energy, $\Omega$, at finite temperature,
both in the semiclassical region and in the dual monopole/dyon theory. In all regions $\Omega$
depends on both the temperature $T$ and the appropriate moduli parameter, and is thus
minimized only for specific values of the moduli parameter, in contrast
to the $T=0$ case where the energy vanishes all over the moduli space. 
Within the validity of perturbation theory, we find that the finite temperature 
Yang--Mills theory 
is stable only at definite points in the moduli space, i.e.
for a specific value of the monopole/dyon mass or when the scalar field expectation
value goes to infinity.  

\vfill
\eject
\pagestyle{plain}

\newcommand \streck {\overline} 
\newcommand \zbar {\bar{z}}
\newcommand \psibar {\overline\psi}
\newcommand \thetaf [4] {\Theta\left[\begin{array}{c}#1 \\ #2
\end{array}\right] (#3 , #4) }
\newcommand \Dsl {D\!\!\!\!/}
\newcommand \dsl {{\partial}\!\!\!\!/}
\setlength{\baselineskip}{23pt}

\noindent{\bf 1. Introduction }
\renewcommand{\theequation}{1.\arabic{equation}}
\setcounter{equation}{0}
\vskip 5mm

In the last few years, our knowledge about
the behavior of supersymmetric non--abelian gauge theories has increased dramatically, both
for $N=1$ and $N=2$ supersymmetry (SUSY) \cite{n1seiberg, seiwitt}. One of the main
achievements has been the derivation of the low energy dynamics, expressed in terms of a 
different, but equivalent, theory. Under certain circumstances this makes it possible to describe 
the original strongly coupled theory by a dual, weakly coupled theory and provides a natural
generalization of Dirac's ``electric--magnetic duality''. 

The classical $N=2$ SU(2) gauge theory has a continuos manifold of inequivalent ground
states. 
This classical moduli space is parametrized by the gauge invariant quantity $u = {\rm Tr} \phi^2 \propto a^2$, and
for $a \neq 0$ the original gauge symmetry is broken down to U(1).
Although there are no quantum corrections that lift the degeneracy among the ground states, 
the quantum theory
is still quite different from the classical one. At large expectation values the theory
is weakly coupled due to asymptotic freedom, and the low energy degrees of freedom are just
the massless excitations of the microscopic theory. Naively, one would 
expect an enhancement of massless fields at $a=0$,
where the full symmetry is restored. However, in the quantum theory $a \neq 0$ all over the moduli space,
and thus the symmetry group is never enlarged. Instead there exist points where 
massive monopoles/dyons become massless \cite{seiwitt}. Thus, at different points in moduli space
it becomes necessary to use different theories to describe the low energy properties. 

Duality in this context is not a symmetry of the theory but rather a way to describe the 
dynamics at a strong electrical 
coupling in terms of e.g. a weakly interacting, magnetically charged theory, where the monopole appears as a
fundamental particle. 

Although the $N=2$ SU(2) gauge theory is not directly relevant for phenomenological studies, it
provides a valuable theoretical framework where nontrivial issues can be tested;
in that respect the supersymmetric theory is worthwhile to study and use as a toy model.
For example, being an asymptotically free, four--dimensional 
nonabelian gauge theory, several of its fundamental properties, such as confinement via condensation of magnetic 
monopoles \cite{seiwitt} and chiral symmetry breaking \cite{seiwitt2},
are also found in QCD. 

Another interesting aspect of field theories (including supersymmetric ones), is their decisive role in
determining the evolution of the early universe, that is their behavior at finite temperature.
Previous studies of the finite temperature effects in SUSY theories have focused on e.g. 
certain fermion condensates in the non--linear SUSY $\sigma$--model 
\cite{jgjw} and symmetry non--restoration effects in softly broken
SUSY theories at high temperature \cite{dvali}.  
Apart from being interesting in its own right, it is not inconceivable that the study of supersymmetric
theories at finite temperature could provide additional qualitative information of a more general nature.  
Therefore, it is of interest to study also the finite temperature effects on 
supersymmetric theories. More specifically, since the 
$N=2$ SUSY Yang--Mills theory has a very rich and complicated structure at $T=0$, that nevertheless is rather well 
understood, one could hope that this theory provides some nontrivial information at finite $T$ as well.

At finite temperature, several properties of a theory may change dramatically from the $T=0$ case.
In particular, a preferred nontrivial minimum of the effective potential can be altered as the temperature is raised.
In ordinary, non--supersymmetric gauge theories this is a well known feature, and in e.g. the
electroweak standard model the SU(2) $\times$ U(1) symmetry is believed to be restored at a
critical temperature $T_c \simeq 200$ GeV \cite{linde}.  
In the $N=2$ SU(2) Yang--Mills theory, the existence of a manifold of ground states 
depends crucially on a delicate balance between the perturbative effects from fermions and bosons, as 
expressed in the non--renormalization theorem \cite{grisaru}. However, at finite temperature the bosons
obey Bose--Einstein statistics whereas fermions obey Fermi--Dirac statistics. Thus, the cancellations
due to supersymmetry do not in general apply to the temperature dependent parts and one
can not {\em a priori} expect the finite $T$ effective potential to remain unchanged. 

As the SUSY Yang--Mills theory is asymptotically free, one would expect the high temperature phase to consist of a
weakly interacting gas of ``electrical'' particles. Another, more interesting 
prospect is the changes in the flatness condition that may occur already at {\em low} temperatures.
A crucial observation regarding the low temperature effects is 
that the flatness of the potential is sensitive to {\em any} disturbance.
In contrast to most other theories, e.g. the electroweak standard model, 
this means that the stable points of the theory may
be altered completely at any temperature $T>0$. Thus even a very small heating of the system is capable of
driving the theory to some specific minima in the moduli space. An interesting consequence is that if
the temperature initially is very high and the ``electric'' plasma phase is cooled, 
then eventually the system would be forced to one of these minima.    
Note that since different minima correspond to a completely different behavior at low energy , once a minima is chosen
the corresponding dynamics is fixed and persists even at very low temperatures. 
This paper is devoted to a study of the existence of such low temperature minima.

To study the low temperature effects on the theory, we will use the low energy effective action
and calculate the free energy perturbatively. To be able to trust perturbation theory the omitted 
corrections of course have to be small. Imposing this condition we will calculate 
the free energy both in the semiclassical
region (large $\left | a \right |$) where the original ``electric'' coupling is small, and in the 
vicinity of the points at which the monopole/dyon becomes massless, where the dual ``magnetic'' 
coupling is small. At a given low temperature, the free energy is then minimized with respect to
the appropriate moduli parameter and the stable points can be determined.

The paper is organized as follows. In the next section we recall some of the main results obtained for the
low energy $N=2$ SUSY SU(2) Yang--Mills theory. Section 3 is devoted to a calculation of the free energy
in the semiclassical region and in section 4 we analyze the thermal effects 
in terms of the dual variables,
around the monopole singularity. Our conclusions and comments on the results are given in section 5,
while our conventions and some technical details are referred to the appendix.   

\vskip 1cm
\noindent{\bf 2. $N=2$ SUSY SU(2) Yang--Mills theory at low energy and $T=0$}
\renewcommand{\theequation}{2.\arabic{equation}}
\setcounter{equation}{0}
\vskip 5mm
    
In this section we set up the model and recapitulate the main results at zero temperature
\cite{seiwitt}. For
a comprehensive review, see e.g. \cite{alvarezhassan, ketov, vecchia}.

The $N=2$ SUSY SU(2) Yang--Mills theory without matter fields consists of a  
chiral $N=2$ field fulfilling a reality condition \cite{n=2field}. In the $N=1$ superspace formulation,
the action can be written as,
\be
S_{{\rm YM}} = \frac{1}{4 \pi} {\rm Im} {\rm Tr} \int d^4 \! x \, \tau \left [ 
\int d^2 \! \theta \, W^{\alpha}
W_{\alpha} \, + \, 2 \int d^2 \! \theta d^2 \! \streck{\theta} \, \streck{\Phi} e^{-2V} \Phi \right ] \ ,
\label{origaction}
\ee 
where $W^{\alpha}$ is a vector multiplet containing a vector boson $A_{\mu}$ and a Weyl fermion
$\lambda$, and the chiral multiplet $\Phi$ contains a complex scalar $\phi$ and a Weyl fermion $\chi$.
The trace is taken over the internal gauge indices, with all fields in the adjoint
representation and the generators normalized according to
${\rm Tr} (T^a T^b ) = \delta^{ab} /2$. The 
function $\tau$ is given by a combination of the coupling constant $g$ and the theta--angle $\theta$,
\be
\tau = \frac{\theta}{2 \pi} +\frac{4\pi i}{g^2} \ .
\ee
This theory is asymptotically free, and at the level of perturbation theory the $\beta$--function
is determined completely by the one--loop contribution \cite{grisaru, howe},
\be
\beta (g) = -\frac{1}{4 \pi^2} g^3 \ . \label{nonabelianbeta}
\ee 
As in QCD, dimensional transmutation takes place and introduces a scale parameter 
$\Lambda$ that governs where the coupling becomes strong.

Writing the classical action (\ref{origaction}) in components and eliminating the auxiliary fields,
the action $S_{{\rm YM}}$ contains a potential for the scalar fields,
\be
V ( \phi^{\dagger}, \phi ) = \frac{1}{2 g^2} {\rm Tr} \left (  [ \phi^{\dagger}, 
\phi ]^{2} \right ) \ .
\ee
This quartic potential is obviously non--negative and is
minimized when $\phi^{\dagger}$ and $\phi$ commute. Up to a
gauge transformation we can take
\be
\phi = \frac{a}{2} \sigma^3 \ ,
\ee
with $a$ an arbitrary complex constant and $\sigma^3$ the usual Pauli matrix. The gauge invariant
quantity that labels the set of inequivalent vacua, the moduli space, in the quantum 
theory is given by
\be 
u = \left \langle {\rm Tr}  ( \phi^2  ) \right \rangle \ .
\ee
For $a \neq 0$ a super--Higgs phenomenon takes place and the SU(2) symmetry is broken down to U(1),
whereas the $N=2$ supersymmetry is still maintained. 

Now consider the low energy part of the spontaneously broken theory, valid for energies
$E \ll \Lambda$.
When $\left | a \right | \gg \Lambda$, the theory is weakly coupled and the massive particles are very heavy
compared to the scale $E$.
As a result, to lowest (zeroth) order in $a^{-1}$ the massive particles can be neglected and the low energy
effective action simply consists of a free massless abelian theory, 
sitting in a single $N=2$ chiral multiplet.
Effective interactions between the massless particles are suppressed by powers of $a$
and appear when the effects of the heavy modes are taken into account.
Actually, all terms up to two derivatives or at most four fermions can be determined completely
(for a proper way of counting in a momentum expansion of a supersymmetric theory, see
\cite{henningson}). In the $N=1$ formalism, the most general U(1) gauge invariant
effective action up to two derivatives, and compatible with $N=2$ supersymmetry is,
\be
S_{{\rm eff}} = {\rm Im} \, \frac{1}{8\pi} \int d^4 \! x \, \left [ \int d^2 \! \theta F^{''} W^{\alpha}
W_{\alpha} \, + \, 2 \int d^2 \! \theta d^2 \! \streck{\theta} \, \streck{\Phi} F^{'} \right ] \ ,
\label{effaction1}
\ee  
where $F(\Phi)$ is the so called 
prepotential and $F^{'} = \del F / \del \Phi$, $F^{''} = \del^2 F / \del \Phi^2$.
The unknown function $F(\Phi)$ is then fixed uniquely at the perturbative level by combining the
non--renormalization theorem \cite{grisaru} and the requirement that 
the effective Lagrangian reflects the U(1)$_R$ anomaly. Together with the nonperturbative instanton
corrections, this gives
\be
F(\Phi) = \frac{i}{2\pi} \Phi^2 \log \left [ \frac{\Phi^2}{\Lambda^2} 
\right ] +\Phi^2 \sum_{n=1}^{\infty} c_n \left ( \frac{\Lambda^2}{\Phi^2} \right )^{2n} \ , 
\ee 
where $c_n$ are coefficients that can be determined from the exact treatment of the problem 
\cite{seiwitt, lerche} or by direct instanton calculations \cite{inst}.

Having derived the low energy action in the large $|a|$ region, one now has to find a description
involving the correct massless fields that is valid when the coupling becomes strong.
In the classical theory, the point $a=0$ corresponds to an enhancement of the symmetry, U(1) 
$\rightarrow$ SU(2), and all the fields become massless. In the quantum theory, a single  
point where extra massless fields appear is not admissible, because in that case
$a^2$ would be a global parametrization of the moduli space, which it cannot be \cite{seiwitt}.
Consequently, there has to be at
least two singularities at $u\neq 0$ where additional particles become massless.
Moreover, the appearance of
massive vector bosons becoming massless at certain points leads to a conformally invariant theory 
in the infrared limit, which contradicts $u \neq 0$.
Therefore, Seiberg and Witten \cite{seiwitt} conjectured that 
there should be exactly two singularities at $\pm \Lambda^2$
in the strong coupling region, and that these two points should correspond to a magnetic monopole    
(at $+\Lambda^2$) or a dyon (at $-\Lambda^2$) becoming massless. 
The dyon has electric and magnetic charge $(n_e,n_m)= \pm (1,-1)$,
while the charge of the monopole is $(n_e,n_m)= \pm (0,1)$.
The correct degrees of freedom in the low energy action hence include the massless fields
that couple locally to the monopole or dyon, together with the effective interactions that results from
integrating out the light soliton. 
This conjecture leads to a consistent picture
and has been justified further later on (see e.g. \cite{ferrari}).

The solitons that become massless at $\pm \Lambda^2$ exist already at the semiclassical level, since 
the original spontaneously broken SU(2) theory admits soliton solutions. 
These solitons fall into a short multiplet and their mass is 
given by the BPS--condition,
\be
M = \sqrt{2} \left | Z \right | \ , \label{monopolemass}
\ee
where $Z$ is the central charge of the supersymmetry algebra. Now, since the relation between the
central charge and the mass
is protected from quantum effects, the mass formula 
(\ref{monopolemass}) remains valid even in the full quantum theory. The central charge can be calculated
explicitly \cite{olive}, with a resulting mass formula for the BPS states,
\be
M = \sqrt{2} \left | an_e + a_D n_m \right | \ .
\ee
Here $a_D = \del F / \del a$ is the variable dual to $a$. In other words, the dual theory in the
vicinity of $u = +\Lambda^2$ is a
magnetic version of $N=2$ QED, where the light monopole has a mass $M_m = \sqrt{2}| a_D|$ and couples
to the dual photon. When $u \rightarrow \Lambda^2$ we have $a_D \rightarrow 0$ and the monopole becomes
massless. The other singularity, where the dyon becomes massless, is related to the
first by a discrete $Z_2$ symmetry that originates from a nonanomalous subgroup of U(1)$_R$. At the point 
$u = -\Lambda^2$
the dyons are massless, corresponding to $(a -a_D) \rightarrow 0$.

Integrating out the monopoles, the resulting effective theory is reminiscent of  
(\ref{effaction1}), but in terms of the dual fields $W^{\alpha}_D$ and $\Phi_D$. Given the 
$\beta$--function for $N=2$ SUSY QED, 
\be
\beta_D = \frac{1}{8\pi^2} g^2_D \ , \label{qedbeta}
\ee
the dual prepotential becomes
\be
F_D(\Phi_D) = -\frac{i}{4\pi} \Phi_D^2 \log \left [ \frac{\Phi_D^2}{\Lambda^2} 
\right ] +\Lambda^2 \sum_{n=1}^{\infty} c^D_n \left ( \frac{\Phi_D}{\Lambda} \right )^{n} \ , 
\label{dualpre}
\ee 
where the coefficients $c^D_n$ can be calculated from the exact solution \cite{lerche}.

\vskip 1cm
\noindent{\bf 3. The free energy in the semiclassical region}
\renewcommand{\theequation}{3.\arabic{equation}}
\setcounter{equation}{0}
\vskip 5mm

In this section we will calculate the free energy for large $|a|$, based on the low energy
effective action.

As has already been noted, for $|a| \gg \Lambda$ the effective action at energy scales $E \ll \Lambda$ 
contains only the remaining massless
fields. These fields are not charged with respect to the unbroken U(1) and are thus free
to first order. All
possible interactions between the massless fields can be expanded in
local, effective vertices and the interactions are suppressed by
powers of the masses of the heavy particles, which are proportional to $|a|$. 
Also, the coupling constant is small due to asymptotic freedom which suppresses the higher order
terms further, albeit only logarithmically.

If the temperature is low, i.e. $T\ll |a|$, the probability of having a real massive particle in
the heat bath is suppressed by the Boltzmann factor $e^{-|a|/T}\ll 1$. Hence we can safely neglect the thermally 
excited massive particles, and need only to take into account the exchange of virtual, massive 
particles between the massless and otherwise free ones.  
Therefore, the dominating 
contributions to the free energy come from the massless particles and their effective interactions.
Furthermore, when the temperature is low we do not expect any large logarithms 
involving the temperature, so $g = g(|a|) \ll 1$. 

The component form of the effective action can be obtained from equation (\ref{effaction1}) in a straightforward
manner.
Eliminating the auxiliary fields and including all vertices that
contribute to the free energy up to $|a|^{-2}$, the action is given by,
\be
S^{(1)}_{{\rm eff}} \hspace{-0.3cm} &=& \hspace{-0.3cm} \int d^4 \! x\, 
\left \{ (\del_{\mu} \rho^{\dagger} )(\del^{\mu} \! \rho) -
\frac{i}{2} \streck{\chi} \dels \chi  -\frac{i}{2} \streck{\lambda} \dels \lambda -
\frac{1}{4} F^{\mu \nu} F_{\mu \nu} + \frac{ig^3}{16 \pi^2}
\left [ \frac{\rho^{\dagger}}{a^{\dagger}} -\frac{\rho }{a} \right ]
F^{\mu \nu} \tilde{F}_{\mu \nu} +
\right.  \nonumber  \\ &+& \hspace{-0.3cm} \left. 
\frac{g^3}{4 \pi^2}\left [ \frac{\rho^{\dagger}}{a^{\dagger}}
+ \frac{\rho}{a}\right ] \left [ (\del_{\mu} \rho^{\dagger} )(\del^{\mu} \!
\rho) -\frac{i}{2} \streck{\chi} \dels \chi  -\frac{i}{2} \streck{\lambda} \dels \lambda -
\frac{1}{4} F^{\mu \nu} F_{\mu \nu} \right ] - \right. \nonumber \\ &-& \hspace{-0.3cm} 
\left. \frac{g^3}{16\pi^2 a} \left [i\streck{\chi} \gamma^{\mu} (1+\gamma_5)\chi (\del_{\mu}\rho ) + 
i \streck{\lambda} \gamma^{\mu} (1+\gamma_5)\lambda (\del_{\mu}\rho ) - \sqrt{2} \, \streck{\lambda}
\Sigma^{\mu \nu}(1+\gamma_5) \chi F_{\mu \nu} \right ] \nonumber + \right. \\ &+& \hspace{-0.3cm} \left. 
\frac{g^3}{16\pi^2 a^{\dagger}} 
\left [i\streck{\chi} \gamma^{\mu} (1+\gamma_5)\chi (\del_{\mu}\rho^{\dagger} ) +
i \streck{\lambda} \gamma^{\mu} (1+\gamma_5)\lambda (\del_{\mu}\rho^{\dagger} ) 
- \sqrt{2} \, \streck{\chi}
\Sigma^{\mu \nu}(1-\gamma_5) \lambda F_{\mu \nu} \right ] - \right. \nonumber \\ &-& \hspace{-0.3cm} 
\left.  
\frac{g^4}{64\pi^2} \left [ \left ( \frac{1}{a^2} \right ) \left \{ \streck{\chi} (1+\gamma_5) \chi \,
\, \streck{\lambda} (1+\gamma_5 ) \lambda \right \} + \left ( \frac{1}{a^{\dagger 2}} \right ) 
\left \{ \streck{\chi} (1-\gamma_5) \chi \,
\, \streck{\lambda} (1-\gamma_5 ) \lambda \right \} \right ] - \right. \nonumber \\ &-& 
\hspace{-0.3cm} \left.
\frac{g^6}{(16\pi^2 )^2} \left [ \left (\frac{1}{a^2} \right ) \left \{ \streck{\lambda} (1+\gamma_5 )
\lambda \, \, \streck{\chi} (1+\gamma_5 ) \chi - \streck{\chi} (1+\gamma_5 )
\lambda \, \, \streck{\lambda} (1+\gamma_5 ) \chi
\right \} \right. \right. + \nonumber \\ &+& \hspace{-0.3cm} 
\left. \left. \left ( \frac{1}{a^{\dagger 2}} \right ) 
\left \{ \streck{\lambda} (1-\gamma_5 )
\lambda \, \, \streck{\chi} (1-\gamma_5 ) \chi - \streck{\chi} (1-\gamma_5 )
\lambda \, \, \streck{\lambda} (1-\gamma_5 ) \chi \right \} \right. +\right. \nonumber \\ &+& 
\hspace{-0.3cm} \left. \left. \left ( \frac{1}{|a|^2} \right ) 
\left \{ \streck{\lambda} (1+\gamma_5)\lambda \, \,
\streck{\lambda} (1-\gamma_5)\lambda +\streck{\chi} (1+\gamma_5)\chi \, \,
\streck{\chi} (1-\gamma_5)\chi \right. \right. \right. + \nonumber \\ &+& \hspace{-0.3cm} 
\left. \left. \left.
2\streck{\chi} (1+\gamma_5)\lambda \, \, \streck{\lambda}(1-\gamma_5)
\chi \right \} \frac{}{} \right ]  \right \} \ , \label{effaction2}
\ee 
where the fermions $\lambda$, $\chi$ have been written as four--component Majorana spinors, $\rho$
is a complex scalar and $F_{\mu \nu}$ is the field strength for the vector boson, with $\tilde{F}_{\mu \nu}$
the dual of $F_{\mu \nu}$. To arrive at the action
(\ref{effaction2}) all fields have been rescaled by the common factor $g$. Finally, the instanton
contributions have been neglected, since they are further suppressed by powers of $(\Lambda / |a|)
\ll 1$.

Since the free energy is a static quantity we will use the imaginary time formulation and apply
by now well established techniques for such finite temperature calculations \cite{kapusta, lebellac}.
As is well known, at finite temperature $T$, the Euclidean space is 
restricted to $\vec{x} \in \Re^3$, $0 \leq x_4 \leq \beta =T^{-1}$ and the bosonic (fermionic)
fields are periodic (antiperiodic) under $x_4 \rightarrow x_4 +\beta$.

To first order, we have a gas of free, massless 
particles and the corresponding contribution to the free energy
density $\Omega$ is just the blackbody radiation,
\be
\Omega^{(1)} = -\frac{\pi^2 T^4}{12} \ . \label{freenergy1}
\ee

At two loops, there are several terms of order $|a|^{-2}$ in equation
 (\ref{effaction2}) that could contribute.
However, it is easily seen that
apart from the vertices proportional to either $(\rho^{\dagger} /a^{\dagger})$ or
$(\rho /a )$, the order $|a|^{-2}$ loops from all the other terms
 vanish trivially, since the remaining integrands are either odd or contain an odd number of
$\gamma$--matrices.
Concerning the contribution from the diagrams involving $(\rho^{\dagger} /a^{\dagger})$ and $(\rho /a)$,
they are all of the form depicted in Fig. 1. Since the momentum integrals are ultra--violet divergent,
they have to be regularized. 
Employing a regularization method that respects the supersymmetry,
such as dimensional reduction\footnote{The conclusions below are
also true when the dimensional regularization method is used.} \cite{siegel, capper}, and using the method outlined
in \cite{thermon=1} to factorize the integrals, the result of the derivative interactions
can be written in the form,
\be
\Omega^{(2)} \hspace{-0.3cm} &\propto & \hspace{-0.3cm}
\left [ \mu^{\epsilon} \int \frac{d^d \! k}{(2\pi)^d} \, \pm \int \frac{d^3 \! k}{(2\pi)^3} 
\int_{-i\infty +\epsilon}^{i\infty +\epsilon} \frac{d k_0}{2\pi i} \frac{1}{e^{\beta k_0} \mp 1} \right ]
\, \times \nonumber \\ &\times & \hspace{-0.3cm}
 \left [ \mu^{\epsilon}\int \frac{d^d \! p}{(2\pi)^d} \frac{1}{p^2} \, + \int \frac{d^3 \! p}{(2\pi)^3} 
\int_{-i\infty +\epsilon}^{i\infty +\epsilon} \frac{d p_0}{2\pi i} \left (\frac{1}{p^2} \right ) \,  
\frac{1}{e^{\beta p_0} - 1} \right ] \label{vanish} \ .
\ee
We have inserted an arbitrary mass parameter $\mu$ to get the correct dimensions, 
and the sum over the Matsubara frequencies $\omega_n$, $\omega_m$ 
have been written as contour integrals, with 
$k_0 = i\omega_n$, $p_0 = i\omega_m$  and $\beta = T^{-1}$. 
The second bracket in the free energy (\ref{vanish}) is due to 
the scalar field, where
$\omega_m = 2 m \pi T $ and in the first bracket the upper sign holds for bosons and the lower for fermions,
where $\omega_n = (2n +1) \pi T $. Now, in the first bracket the first term vanishes due to the definition
of integration in dimensional regularization. In the second term inside the first bracket
we can close the contour in the right half plane of
the complex $k_0$--plane, and since the integrand has no poles there is no contribution from this term either. 
As a consequence, the expression in equation (\ref{vanish}) vanishes and $\Omega^{(2)} =0$. Thus
there are no $T^6 /|a|^2$ corrections to the free energy.

\begin{center}
\begin{picture}(300,80)(100,40)
\DashArrowLine(108,75)(192,75){4}
\ArrowArc(150,50)(50,30,150)
\ArrowArc(150,100)(50,210,330)
\GCirc(108,75){2}{0.5}
\GCirc(192,75){2}{0.5}
\DashArrowLine(230,100)(260,100){4}
\Text(265,100)[l]{= scalar field propagator}
\ArrowLine(230,75)(260,75)
\Text(265,75)[l]{= scalar, fermion or}
\Text(265,62)[l]{\hspace{0.3cm} vector field propagator}  
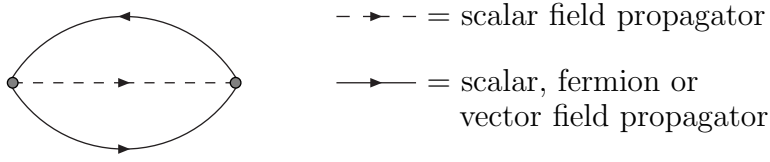
\end{picture} \\ {\sl {\rm Fig. 1.} Nontrivial diagrams in $\Omega^{(2)}$, at order $|a|^{-2}$.}
\end{center}
 
Expanding the action (\ref{effaction1}) further to include higher orders in $|a|^{-1}$, 
it is easy to see that there
are no loop corrections to the blackbody radiation that are suppressed by $a^3$. 
Therefore, the leading correction
term to the free part in (\ref{freenergy1}) has to be at least of order
{\cal O}$(T^8 /|a|^4)$. These terms will also
be of order $g^8$ or higher in the coupling constant.

However, the order $T^8 /|a|^{4}$--terms do not only originate from an expansion of the effective
action in (\ref{effaction1}). Apart from this holomorphic piece, 
there are also non--holomorphic contributions that start at the level of
four--derivative terms \cite{nonhol1, nonhol2, nonhol3}. 
In the $N=1$ superspace formulation
the non--holomorphic part of the effective action is \cite{henningson},
\be
S_{{\rm eff}}^{{\rm non.hol.}} \hspace{-0.3cm} &=& \hspace{-0.3cm}
-\frac{2}{(16\pi)^2} \int d^4\! x \, d^2 \! \theta \, d^2 \! \streck{\theta} \,
\left[ K_{\Phi \streck{\Phi}} \left \{ D^{\alpha} D_{\alpha} \Phi \streck{D}_{\dot{\alpha}} \
\streck{D}^{\dot{\alpha}} \streck{\Phi} +2\streck{D}_{\dot{\alpha}}D^{\alpha}\Phi D_{\alpha}
\streck{D}^{\dot{\alpha}}\streck{\Phi} + \right. \right. \nonumber \\ &+& 
\hspace{-0.3cm} \left. \left.
4D^{\alpha} W_{\alpha} \streck{D}_{\dot{\alpha}} 
\streck{W}^{\dot{\alpha}}  -4D^{(\alpha} W^{\beta )}D_{(\alpha} W_{\beta )}
-4\streck{D}_{(\dot{\alpha}} \streck{W}_{\dot{\beta} )}\streck{D}^{(\dot{\alpha}} 
\streck{W}^{\dot{\beta} )} -2D^{\alpha}D_{\alpha} (W^{\beta}W_{\beta} ) - 
\right. \right. \nonumber \\ &-& \hspace{-0.3cm}
\left. \left. 2\streck{D}_{\dot{\alpha}}\streck{D}^{\dot{\alpha}} 
( \streck{W}_{\dot{\beta}}
\streck{W}^{\dot{\beta}} ) \right \} -2 K_{\Phi \Phi \streck{\Phi}} W^{\alpha} W_{\alpha} D^{\beta}
D_{\beta} \Phi  - 2 K_{\Phi \streck{\Phi} \streck{\Phi}} \streck{W}_{\dot{\alpha}} 
\streck{W}^{\dot{\alpha}}  \streck{D}_{\dot{\beta}}\streck{D}^{\dot{\beta}} \Phi^{\dagger}  +
\right. \nonumber \\ &+& \hspace{-0.3cm}
\left.  K_{\Phi \Phi \streck{\Phi} \streck{\Phi}} \left \{ 
4W^{\alpha} W_{\alpha} \streck{W}_{\dot{\alpha}} \streck{W}^{\dot{\alpha}} -8W^{\alpha}D_{\alpha} \Phi
\streck{W}_{\dot{\alpha}} \streck{D}^{\dot{\alpha}} \Phi^{\dagger} \right \} \right ] 
\label{nonholaction} \ ,
\ee
where we have defined
\be
K= \log \left ( \frac{\Phi}{a} \right ) \log \left (\frac{\streck{\Phi}}{a^{\dagger}} \right ) \ ,
\ee
and the subscripts on $K$ denote derivatives with respect to that argument.  
Rewriting equation 
(\ref{nonholaction}) in component form and taking into account only the terms up to $|a|^{-4}$,
we find that all possible divergent terms vanish one by
one when the dimensional reduction, or dimensional regularization, procedure is used.
The remaining terms in the effective Lagrangian that 
contribute to the finite temperature effects are 
given by, in the Weyl two--component notation,
\be
L^{(2)}_{{\rm eff}} \hspace{-0.3cm} &=& \hspace{-0.3cm}
\frac{-g^4}{16 \pi^2 |a|^4} \left [ 2(\del_{\mu} \rho^{\dagger})(\del^{\mu} \!
\rho )(\del_{\nu} \rho^{\dagger})(\del^{\nu} \! \rho ) +2\rho^{\dagger} \rho (\del_{\mu}\del_{\nu}
\rho^{\dagger} )(\del^{\mu}\del^{\nu} \! \rho ) + 2\left (\frac{F^{\mu \nu}F_{\mu \nu}}{4} \right )^2
 + \right. \nonumber \\ &+& \hspace{-0.3cm}
\left. 2\left (\frac{F^{\mu \nu}\tilde{F}_{\mu \nu}}{4} \right )^2
+ \left \{ 2(\streck{\chi}\, \streck{\sigma}^{\mu} \del_{\mu}\chi )
(\del_{\nu}\streck{\chi}\, \streck{\sigma}^{\nu} \chi )  +
2(\chi \del_{\mu} \chi )(\streck{\chi} \del^{\mu} \streck{\chi}) +(\chi \rightarrow \lambda )
\right \}  + \right. \nonumber \\ &+& \hspace{-0.3cm}
\left. \left \{ i(\del^{\nu}\streck{\chi}\, \streck{\sigma}^{\mu} 
\chi )(\del_{\mu} \rho^{\dagger} )(\del_{\nu} \rho ) -i(\streck{\chi} \, \streck{\sigma}^{\mu}\del^{\nu}
\chi )(\del_{\nu} \rho^{\dagger} )(\del_{\mu} \rho ) +i\rho (\del_{\mu} \del_{\nu} 
\rho^{\dagger} )(\streck{\chi} \, \streck{\sigma}^{\nu} \del^{\mu} \chi ) - \right. \right. \nonumber \\
&-& \hspace{-0.3cm} \left. \left. i\rho^{\dagger} (\del_{\mu} \del_{\nu} \rho )(\del^{\mu}\streck{\chi} 
\, \streck{\sigma}^{\nu} \chi ) + i(\streck{\chi}\, \streck{\sigma}^{\mu}\del_{\nu}\chi )
F_{\mu \delta}F^{\nu \delta} - i(\del_{\nu}\streck{\chi}\, \streck{\sigma}^{\mu}\chi )
F_{\mu \delta}F^{\nu \delta} +
(\chi \rightarrow \lambda) \right \} - \right. \nonumber \\ &-& \hspace{-0.3cm}
\left. 4(\del_{\mu}\rho^{\dagger})(
\del^{\nu} \! \rho )F^{\mu \delta}F_{\nu \delta} - (\lambda \sigma^{\mu}\streck{\sigma}^{\nu}
\del_{\nu}\chi )(\streck{\chi}\del_{\mu}\streck{\lambda} ) + (\chi\del_{\mu}\lambda )
(\streck{\lambda}\streck{\sigma}^{\nu}
\sigma^{\mu} \del_{\nu}\streck{\chi}) + \right. \nonumber \\
&+& \hspace{-0.3cm} \left. (\lambda\sigma^{\nu}\streck{\sigma}^{\mu}\del_{\nu}\chi )
(\streck{\chi}\del_{\mu}\streck{\lambda}) -(\chi\del_{\nu}\lambda )(\streck{\lambda} \streck
{\sigma}^{\nu}\sigma^{\mu} \del_{\mu}\chi ) \left. \frac{}{} \right.
\right ] +{\cal {\rm O}} \left (g^6/|a|^4 \right ) \ , \label{nonholcomp}
\ee
where all the fields have been rescaled with a factor $g$, in accordance with conventions used 
for the holomorphic action (\ref{effaction2}). The terms of order $g^6$ appear when the lowest 
order equations of motion for the auxiliary fields, derived from the holomorphic action  (\ref{effaction1}), 
are substituted back into the action (\ref{nonholcomp}).
Since the lowest order terms in the
non--holomorphic part are of order $g^4$, it is clear that the leading correction terms
to the free--field result come from equation (\ref{nonholcomp}), and not from the holomorphic part (\ref{effaction2}).
 
The contributions to the free energy
come from two--loop diagrams of the form shown in Fig. 2, where each loop, independently of the
other, is any of the bosons ($\rho$, $A_{\mu}$) or fermions ($\lambda$, $\chi$).

\begin{center}
\begin{picture}(300,80)(0,0)
\ArrowArc(120,35)(30,0,360)
\DashArrowArc(180,35)(30,180,540){4}
\GCirc(150,35){3}{0.5}
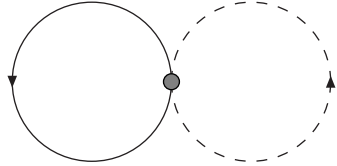
\end{picture} \\ {\sl {\rm Fig. 2.} Two--loop contribution in $\Omega^{(3)}$, at order $|a|^{-4}$.}
\end{center}
Evaluating the different diagrams, we find the following contributions,
\be
&&\bullet \hspace{0.2cm} \mbox{fermion-fermion terms:} \hspace{0.5cm} 
\Omega^{(3)}_{{\rm ff}} = \frac{49\pi^2 g^4T^8}{43200|a|^4} \ , \nonumber \\
&&\bullet \hspace{0.2cm} \mbox{scalar-scalar terms:} \hspace{0.5cm} \hspace{18pt}
\Omega^{(3)}_{{\rm ss}} = \frac{\pi^2 g^4T^8}{2700|a|^4} \ , \nonumber \\
&&\bullet \hspace{0.2cm} \mbox{vector-vector terms:} \hspace{0.5cm} \hspace{15pt}
\Omega^{(3)}_{{\rm vv}} = \frac{\pi^2 g^4T^8}{2700|a|^4} \ , \nonumber \\
&&\bullet \hspace{0.2cm} \mbox{scalar-fermion terms:} \hspace{0.5cm} \hspace{9pt}
\Omega^{(3)}_{{\rm sf}} = \frac{7\pi^2 g^4T^8}{5400|a|^4} \ , \nonumber \\
&&\bullet \hspace{0.2cm} \mbox{vector-fermion terms:} \hspace{0.5cm} \hspace{8pt}
\Omega^{(3)}_{{\rm vf}} = \frac{7\pi^2 g^4T^8}{5400|a|^4} \ ,  \nonumber \\
&&\bullet \hspace{0.2cm} \mbox{vector-scalar terms:} \hspace{0.5cm} \hspace{17pt}
\Omega^{(3)}_{{\rm vs}} = \frac{\pi^2 g^4T^8}{1350|a|^4} \ . \label{freenergy2}
\ee
As expected, since the physical degrees of freedom for the complex scalar field
and the vector boson are the same, their contributions are identical. 
Also, the fermions and bosons obey different statistics, and the result of having a fermion loop
or a boson loop should not be equal. However, as is apparent from the terms in (\ref{freenergy2}), they all give
results that add up instead of tending to cancel each other, as is the case in SUSY theories at $T=0$.
This additivity of the thermal effects is in accordance with earlier observations on the finite $T$ behavior
of SUSY theories \cite{das}.
  
Combining the free--field part in (\ref{freenergy1}) and adding all the terms from 
(\ref{freenergy2}), we finally have 
\be
\Omega = -\frac{\pi^2 T^4}{12} + \frac{g^4\pi^2T^8}{192|a|^4} +{\cal {\rm O}} \! \! 
\left ( \frac{g^6T^8}{|a|^4} \right ) \ . \label{energyrunaway}
\ee 
Since we have found a dependence on both $T$ and $|a|$ in $\Omega$, we can now regard the temperature
as being a fixed external parameter and minimize the free energy with respect to $|a|$.
As is clear from equation (\ref{energyrunaway}),
the free energy decreases as $|a|$ becomes larger, and since a larger $|a|$ corresponds
to a smaller coupling constant, the approximations made in (\ref{energyrunaway}) become
even better. Thus, the prediction from the perturbative calculation is reliable and 
the free energy tends towards its minimum value when $|a| \rightarrow \infty$. 
This run--away solution in the semiclassical region is somewhat 
reminiscent of the superpotential solution in
$N=1$ SU(N$_c$) Yang--Mills theories with N$_f$ (N$_f$ $<$ N$_c$) flavors
\cite{runawayn=1}, although the origin of the behavior is of course
completely different. It is a rather interesting result that the semiclassical region is well
defined at $T=0$, but tends to an unphysical solution for any $T >0$. 

However, it should be stressed that the above conclusion is only trustworthy to the extent that higher 
order corrections in $g$ can be neglected, and is strictly speaking only true in the far asymptotic region.
In fact, if the order $g^6$
contribution from (\ref{nonholcomp}) 
and the order $g^8$ correction from the holomorphic part (\ref{effaction2}) are non--zero, and if the numerical
coefficient in front is large, one can not exclude that the non--asymptotic region has another behavior, with
an additional stable point at large, but finite $|a|$. 

\vskip 1cm
\noindent
{\bf 4. The free energy in the dual theory}
\renewcommand{\theequation}{4.\arabic{equation}}
\setcounter{equation}{0}
\vskip .5cm

In this section we will use the dual formulation of the theory and calculate the free
energy around the monopole singularity.

As $|a|$ decreases, the theory moves away from the semiclassical region and the system 
becomes strongly coupled. Although the condition $|a| \gg T$ might be fulfilled,
the strong coupling  renders a perturbative calculation in the original ``electric''
variables meaningless.
Fortunately, the powerful tool of duality provides a mean to perform the calculation
in certain regions of the strong coupling regime. When entering the strong coupling region
the lightest BPS--states consist of the $\pm (0,1)$ monopoles  around $u = \Lambda^2$ and
the $\pm (1,-1)$ dyons for $u \simeq - \Lambda^2$.
 
As stated earlier, around the point $u = \Lambda^2$ the low energy theory is a magnetic 
version of $N=2$ QED, where the magnetic monopoles have a mass $M_m = \sqrt{2} |a_D|$
and become massless as $u \rightarrow \Lambda^2$. The dual coupling constant $g_D$ is small as
long as the characteristic energy scale $E$ is small, $E \ll \Lambda $. 

Restricting the calculation to $T \ll |a|$ and $T \ll \Lambda$ still permits an arbitrary relation between 
the temperature and the monopole mass. Two extreme limits can be calculated analytically, namely $T \gg M_m$ and
$T \ll M_m$. In the first limit monopoles easily become thermally excited, whereas in the latter
one has to use an effective action where the monopoles have been integrated out. 

Now, consider first the case when $|a_D|\ll T$ (and $T \ll |a|$, $T \ll \Lambda$).
Under these conditions we are in the  weak coupling region, and the heat bath consists of both
the massless $N=2$ chiral multiplet (where the ``dual photon'' is the vector boson) and the
massive but light $N=2$ monopole hypermultiplet. At this temperature the contributions
 from any other particles, such the dyons, 
are neglectable since their mass is proportional to $|a|$. Thus we are in a high temperature
phase with respect to the light degrees of freedom.

Using $N=1$ superspace, the Lagrangian of the theory is,
\be
L_{{\rm eff}}^{(D)} \hspace{-0.3cm} &=& \hspace{-0.3cm}
\frac{1}{8\pi} {\rm Im} \left \{ \tau_D \left [ \int d^2 \! \theta \,
W^{\alpha}_D W_{D \alpha} + 2\int d^2 \! \theta d^2 \! \streck{\theta}\, \streck{\Phi}_D
\Phi_D \right ]
\right \} + \int d^2\! \theta d^2\! \streck{\theta} \left [ \streck{Q}_D e^{-2V_D}Q_D  + \right.
\nonumber \\ &+& \hspace{-0.3cm} \left.
\streck{\tilde{Q}}_D e^{2V_D} \tilde{Q}_D \right ] +\left \{ \int d^2\theta \left [ 
\sqrt{2} \tilde{Q}_D \Phi_D Q_D +m\tilde{Q}_D Q_D \right ] + \, {\rm h.c.} \right \} \ , 
\label{monopoleaction} 
\ee   
where $m =\sqrt{2} a_D$. $W_D$, $\Phi_D$ belong to the $N=2$ chiral multiplet and 
$Q_D$, $\tilde{Q}_D$ to the hypermultiplet, where  $Q_D$ and $\tilde{Q}_D$
have opposite gauge quantum numbers.
After some rescalings, the component form of the Lagrangian (\ref{monopoleaction})
becomes,
\be
L_{{\rm eff}}^{(D)} \hspace{-0.3cm} &=& \hspace{-0.3cm}
-\frac{1}{4}F_{\mu \nu}F^{\mu \nu} -\frac{i}{2}\streck{\lambda} \dels
\lambda-\frac{i}{2}\streck{\chi} \dels \chi -i \streck{\psi} \Dels \psi + 
(\del_{\mu}\rho^{\dagger} )(\del^{\mu} \! \rho ) +(D_{\mu}\phi )^{\dagger} (D^{\mu} \! \phi ) +
\nonumber \\ &+& \hspace{-0.3cm} 
(\tilde{D}_{\mu}\tilde{\phi} )^{\dagger} (\tilde{D}^{\mu} \! \tilde{\phi} ) 
+ \frac{ig_D}{\sqrt{2}} \left [ \streck{\psi}(1-\gamma_5)\lambda \phi - \streck{\lambda}
(1+\gamma_5)\psi \phi^{\dagger} - \streck{\lambda}(1-\gamma_5)\psi \tilde{\phi} + \right.
\nonumber \\ &+& \hspace{-0.3cm} 
\left. \streck{\psi}(1+\gamma_5)\lambda \tilde{\phi}^{\dagger} \right ] -\frac{g_D}{\sqrt{2}}
\left [ \rho \streck{\psi}(1+\gamma_5)\psi +\rho^{\dagger} \streck{\psi}(1-\gamma_5)\psi 
+\tilde{\phi} \streck{\chi} (1+\gamma_5)\psi + \right. \nonumber \\ 
&+& \hspace{-0.3cm} 
\left. \tilde{\phi}^{\dagger} \streck{\psi} (1-\gamma_5)\chi +\phi \streck{\psi}(1+\gamma_5)
\chi +\phi^{\dagger} \streck{\chi} (1-\gamma_5)\psi \right ] + \nonumber \\
&+& \hspace{-0.3cm} \sqrt{2}g_D \left [ M_m (\tilde{\phi}^{\dagger}\tilde{\phi} 
+ \phi^{\dagger}\phi )(\rho^{\dagger} +\rho ) - \sqrt{2}g_D \rho^{\dagger}\rho (
\phi^{\dagger}\phi + \tilde{\phi}^{\dagger}\tilde{\phi} ) \right ] - \nonumber \\
&-&\hspace{-0.3cm} \frac{g^2_D}{2} \left (
\phi^{\dagger}\phi + \tilde{\phi}^{\dagger}\tilde{\phi} \right )^2  
+ M_m \streck{\psi} \psi - M^2_m (\phi^{\dagger}\phi + \tilde{\phi}^{\dagger}\tilde{\phi} ) \ ,
\label{dualcomponent}
\ee 
where $\Dels = (\del^{\mu} -ig_DA^{\mu})\gamma_{\mu}$, $D^{\mu} = \del^{\mu} -ig_DA^{\mu}$
and $\tilde{D}^{\mu} = \del^{\mu} +ig_DA^{\mu}$.
The Lagrangian has been written in a four--component form, and 
$(\rho ,A_{\mu},\chi ,\lambda )$ belong to the chiral multiplet whereas
$( \phi ,\tilde{\phi}, \psi )$ come
from the hypermultiplet (and $\psi$ is a Dirac spinor).

As before, the lowest order contribution to the free energy is just the blackbody radiation.
Denoting the free energy density in the dual theory by $\Omega_D$ at $T\gg M_m$, 
the contribution from the free part is,
\be
\Omega^{(1)}_D = -\frac{\pi^2 T^4}{6} +\frac{M_m^2 T^2}{4} -\frac{M_m^3 T}{3\pi}  +{\cal {\rm O}} \! \!
\left ( M_m^4 \log ( M_m / \pi T)\right ) \ , \label{freedualpart} 
\ee
where the free energy for the massive particles have been expanded in powers of $(M_m /T)$. 
As is clear from equation (\ref{freedualpart}), the noninteracting part of the
free energy becomes smaller when $M_m \rightarrow 0$. Based only on the part of the free energy displayed in
(\ref{freedualpart}), one would conclude that $M_m = 0$ represents a minimum. However, as will be clear,
the interaction effects are not insignificant.

Taking into account also the interactions from the Lagrangian (\ref{dualcomponent}),
the next contributions to $\Omega_D$ are two--loop effects, which are suppressed
by ${\cal {\rm O}} (g_D^2)$. Using the method developed in \cite{nortontrick, kapustatrick}, 
their contributions can be calculated in a power series in  $(M_m /T)$. We find,
\be
\bullet && \hspace{-0.5cm} \mbox{fermion-vector terms:} \nonumber \\ && \Omega^{(2)}_{_D \, {\rm fv}} = 
\frac{5g^2_DT^4}{288} \left \{ 1 +\frac{18}{5\pi^2}\left (\frac{M_m^2}{T^2} \right ) \left [
\gamma -\frac{1}{2} +\log \left (\frac{M_m}{\pi T} \right ) \right ] \right \}
\ , \nonumber \\
\bullet && \hspace{-0.5cm} \mbox{scalar-vector terms:} \nonumber \\ && \Omega^{(2)}_{D \, {\rm sv}} = 
\frac{7g^2_DT^4}{144} \left \{ 1 -\frac{24M_m}{7T\pi} -\frac{12}{7\pi^2}\left (\frac{M_m^2}{T^2} \right ) \left [ \gamma -\frac{5}{4} +\log \left (\frac{M_m}{4\pi T} \right ) \right ] \right \}
\ , \nonumber \\
\bullet && \hspace{-0.5cm} \mbox{Yukawa terms:} \nonumber \\ && \Omega^{(2)}_{D \, {\rm y}} = 
\frac{25g^2_DT^4}{288}  \left \{ 1 -\frac{48M_m}{25T\pi} +\frac{72}{25\pi^2}\left (\frac{M_m^2}{T^2} \right ) \left [ \frac{3\gamma}{4} -\frac{3}{8} +\frac{9}{12}\log \left (\frac{M_m}{\pi T} 
\right ) +\frac{1}{3}\log (4) \right ] \right \} , \nonumber \\
\bullet && \hspace{-0.5cm} \mbox{scalar-scalar terms:} \nonumber \\ && \Omega^{(2)}_{D \, {\rm ss}} = 
\frac{7g^2_DT^4}{144} \left \{ 1 -\frac{30M_m}{7T\pi} -\frac{3}{7\pi^2}\left (\frac{M_m^2}{T^2} \right ) \left [ 5\gamma -\frac{23}{2} +5\log \left (\frac{M_m}{4\pi T} \right ) \right ] \right \} \ ,
\label{freedual2}
\ee
where $\gamma$ is Euler's constant and terms of order {\cal O}$(g_D^2 M_m^3T)$ 
and higher have been omitted.

Adding all the different parts in equation (\ref{freedual2}) together with the free--field result in
(\ref{freedualpart}) we have,
\be
\Omega_D \hspace{-0.3cm} &=& \hspace{-0.3cm}
-\frac{\pi^2 T^4}{6} \left [ 1- \frac{29g^2_D}{24\pi^2} \right ] -\frac{13g^2_DM_mT^3}
{24\pi} +\frac{M_m^2T^2}{4} \left [1 +\frac{4g^2_D}{\pi^2} \left \{ \frac{7}{32} +\frac{\gamma}{16}
+ \right. \right. \nonumber \\ &+& \hspace{-0.3cm} \left. \left.
\frac{1}{16}\log \left (\frac{M_m}{\pi T} \right ) +\frac{13}{48} \log (4) \right \} \left. \frac{}{}
\right. \right ] 
-\frac{M_m^3 T}{3\pi} +{\cal {\rm O}} \! \! \left ( g_D^2 M_m^3T, M_m^4 \right ) \ .
\label{totfreedual}
\ee

The total free energy in
(\ref{totfreedual}) displays some remarkable features. When $T\gg M_m$ the relevant
scale is set by the temperature, so we expect that $g_D^2 = g_D^2 (T)$. Now, for a sufficiently small 
$M_m/T$ ratio,
the term proportional to $g_D^2M_mT^3$ is comparable to $M_m^2T^2$.
In other words, the interactions may change the minimum (at $M_m=0$) found from the blackbody radiation only. 
From $\Omega_D$ in (\ref{totfreedual}), the dominant mass dependent 
contribution $\Omega_{M_m}$ in $\Omega_D$ is
\be
\Omega_{M_m} = \left ( \frac{M_m}{4} -\frac{13g_D^2T}{24\pi} \right ) M_m T^2 \ ,
\ee
and therefore the position of the extremum of the free energy is,
\be
M_m = \frac{13g_D^2T}{12\pi}  \ . \label{turnpoint}
\ee
In fact, this point represents the minimum. Hence the free energy is minimized when equation 
(\ref{turnpoint}) is satisfied,
that is when the zero temperature mass is of the same magnitude as the thermally induced ``electric'' mass.
By the $Z_2$ symmetry this implies a similar result to be valid in the region where the dyon
is the lightest massive particle.

We now turn to the opposite extreme in the dual region, i.e. when $M_m \gg T$. We also require
$M_m \ll \Lambda$ to be able to trust the perturbation theory. When the parameters fulfill the above
conditions the thermally excited monopoles only contribute with a factor $e^{-M_m /T}$. 
Of course, the exponentially small term reflects the probability of having real monopoles in the
heat bath. As in the semiclassical region,
it is clear that the dominating effect from the monopoles is due to the virtual exchange of monopoles between the
massless particles, and it is appropriate to use a low energy effective action.

The holomorphic part of the effective action is given by equation
(\ref{effaction1}), but in terms of the dual
variables $W_D$, $\Phi_D$ and with the dual prepotential (\ref{dualpre}).
The contribution from the blackbody radiation to the free energy $\tilde{\Omega}_D$ is, 
\be
\tilde{\Omega}^{(1)}_D = - \frac{\pi^2 T^4}{12} \ . \label{freedualmt}
\ee

As in the previous section, we have to include the interactions to incorporate the mass parameter, 
in this case $|a_D|$.
In the semiclassical region the holomorphic part did not contribute to 
the leading correction terms in the free energy, and this result
persists in the dual theory as well. The reason is simply that the expansion of the dual prepotential
is completely analogous to the case in the semiclassical region, with the 
expansion parameter $a$ in the semiclassical region replaced by $a^n \rightarrow -2a_D^n$, 
except in the coupling constant.  
This in turn is a consequence of the fact that the coefficient in the $\beta$--function of $N=2$ QED 
differs by a factor $-1/2$ from the SU(2) case, as a comparison between 
equation (\ref{nonabelianbeta})
and (\ref{qedbeta}) shows.

However, just as in the case of ordinary, non--supersymmetric QED one should expect higher order
interactions coming from an Euler--Heisenberg type of effective Lagrangian \cite{schwinger}. In fact,
such a supersymmetric extension was discussed in \cite{nonhol2}, and corresponds to nothing but the
non--holomorphic contribution to the effective action. In that sense, the non--holomorphic action in the
semiclassical region is the supersymmetric extension of a nonabelian Euler--Heisenberg action, whereas 
for the magnetic monopoles it is an extension of the abelian case.

As in the holomorphic part, the effective action in the dual region is given by 
replacing $a^n \rightarrow -2a_D^n$ in  (\ref{nonholcomp}) (see
also \cite{nonhol3}). Therefore, the lowest order
correction to the free energy, when $M_m \gg T$, is
\be
\tilde{\Omega}^{(3)}_D = -\frac{g^4_D\pi^2T^8}{384|a_D|^4} +{\cal {\rm O}} \! \!
\left ( \frac{g_D^6T^8}{|a_D|^4}
\right ) \ . \label{nonholdualenergy}
\ee
The total free energy in the dual region, valid for $M_m \gg T$ is then obtained by
combining equation (\ref{freedualmt}) and (\ref{nonholdualenergy}),
\be
\tilde{\Omega}_D  = - \frac{\pi^2 T^2}{12} -\frac{g^4_D\pi^2T^8}{384|a_D|^4} + {\cal {\rm O}} \! \!
\left ( \frac{g_D^6T^8}{|a_D|^4} \right ) \ .
\ee
Minimizing the free energy we thus find that the theory prefers a smaller mass, a statement that of course is 
valid only as long as both $M_m \gg T$ and the omitted higher order corrections are neglectable.
Again this result is transferred to the region around the dyon singularity by
the $Z_2$ symmetry.

To summarize this section, we have found that when $M_m \gg T$ the free energy becomes smaller as the mass
decreases. When the mass is sufficiently small, the theory enters a high temperature region where $T \gg M_m$.
As expected, the mass continues to decrease, but rather unexpectedly only down to $M_m \propto g^2_D T$. 
If the mass initially is smaller, it will increase instead, up to the minimum of the free energy at 
$M_m \propto g^2_D T$. 
Note though, that as in the semiclassical region we can not exclude another 
minimum in a region where higher orders in the
coupling constant become relevant. Since the dual coupling increases as the mass becomes larger, such a minimum
could be present when the theory is sufficiently far away from $u = \Lambda^2$, but still in the regime where the dual
theory is well defined.

Using the $Z_2$--symmetry, we also conclude that
the same situation as described above takes place around the singularity at
$u = -\Lambda^2$, where the dyons are light.

\vskip 1cm
\noindent
{\bf 5. Conclusions}
\renewcommand{\theequation}{5.\arabic{equation}}
\setcounter{equation}{0}
\vskip .5cm

We have calculated the free energy in the low energy $N=2$ supersymmetric SU(2) Yang--Mills theory, using 
the low energy effective action description, and would now like to make some comments and 
remarks about our results. It should be emphasized that some of the conclusions are naturally rather speculative
at the present stage.

In the semiclassical region where $|a| \gg \Lambda$, we find that the lowest order 
correction terms come from the non--holomorphic part
of the effective action, and as a result 
the free energy decreases as the moduli parameter $|a|$ increases. The stable point corresponds to the 
limit $|a| \rightarrow \infty $. Using
the dual description in the strong coupling regime, 
we find that the theory flows towards a stable point where
the monopole mass $M_m$ is of the order of the thermally induced mass, $M_m \propto g^2_D T$.
The discrete $Z_2$--symmetry relates the monopole singularity to
the region where the dyons become the light solitons, and therefore implies 
that the monopole results are valid for the dyons as well.   
Thus the scenario depicted in Fig. 3 emerges, where region III corresponds to the semiclassical region
$|a| \gg \Lambda$, and the two other regions correspond
to a neighborhood where the dual theory is valid;
region II to where the soliton mass $M_s$ satisfies $M_s \gg T$ and region I to $T \gg M_s$.

\begin{center}
\begin{picture}(400,120)(0,0)
\LinAxis(200,0)(200,100)(4,4,1,1,1)
\LinAxis(0,50)(400,50)(4,4,1,1,1)
\Vertex(250,50){1.5}
\DashCArc(250,50)(10,0,360){1.5}
%INNER ARROWS IN I
\LongArrow(250,50)(255,55)
\LongArrow(250,50)(255,45)
\LongArrow(250,50)(245,45)
\LongArrow(250,50)(245,55)
%OUTER ARROWS IN I
\LongArrow(265,65)(260,60)
\LongArrow(265,35)(260,40)
\LongArrow(235,35)(240,40)
\LongArrow(235,65)(240,60)
\Text(238,52)[br]{I}
%ARROWS FOR II
\LongArrow(280,80)(275,75)
\LongArrow(280,20)(275,25)
\LongArrow(220,20)(225,25)
\LongArrow(220,80)(225,75)
\Text(285,52)[bl]{II}
%REGION III
\DashCArc(200,50)(197,-15,15){1.5}
\LongArrow(383,64)(390,65)
\LongArrow(381,79)(388,80)
\LongArrow(383,36)(390,35)
\LongArrow(381,21)(388,20)
\Text(395,52)[br]{III}
%
%REFLECTIONS IN LEFT HALFPLANE
%
\Vertex(150,50){1.5}
\DashCArc(150,50)(10,0,360){1.5}
%INNER ARROWS IN REFL. I
\LongArrow(150,50)(155,55)
\LongArrow(150,50)(155,45)
\LongArrow(150,50)(145,45)
\LongArrow(150,50)(145,55)
%OUTER ARROWS IN REFL. I
\LongArrow(165,65)(160,60)
\LongArrow(165,35)(160,40)
\LongArrow(135,35)(140,40)
\LongArrow(135,65)(140,60)
\Text(163,52)[bl]{I}
%ARROWS FOR REFL. II
\LongArrow(180,80)(175,75)
\LongArrow(180,20)(175,25)
\LongArrow(120,20)(125,25)
\LongArrow(120,80)(125,75)
\Text(115,52)[br]{II}
%REGION REFL. III
\DashCArc(200,50)(197,165,195){1.5}
\LongArrow(17,64)(10,65)
\LongArrow(17,79)(10,80)
\LongArrow(17,36)(10,35)
\LongArrow(17,21)(10,20)
\Text(5,52)[bl]{III}
\Line(370,108)(370,100)
\Line(370,100)(378,100)
\Text(372,102)[bl]{$u$}
\end{picture}
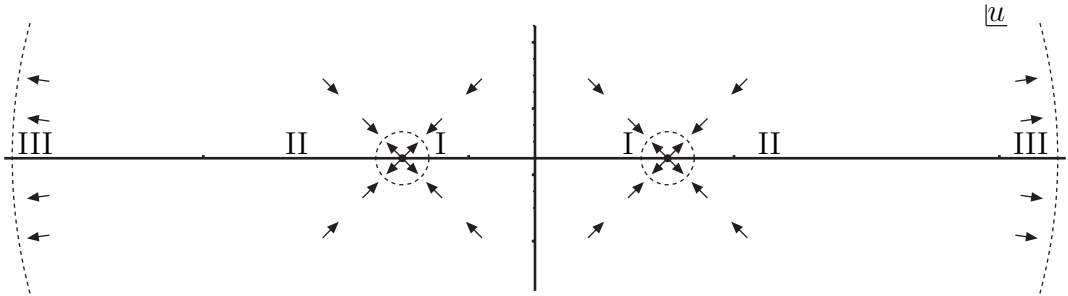
\\ {\sl {\rm Fig. 3.} Directions where the free energy decreases, at different parts in moduli space.
Dotted lines denote the stable minima.}
\end{center}

As mentioned earlier, one can not rule out the existence of other minima, positioned at large but finite $|a|$ and
at a large value of the soliton mass, $M_s$, in the dual theory. In order to
establish whether such minima actually appear, one would have to calculate the logarithmic
corrections coming
from higher orders in the coupling constant, but having the same suppression of powers of mass as the dominant effect
(i.e. $|a|^{-4}$ or $M_s^{-4}$). 
At this stage the presence of such minima  can neither be proved, nor excluded.

In the asymptotic region, where the minimum is reached for
$|a| \rightarrow \infty $, the theory does not seem to be well defined in the finite temperature
case. Nevertheless, the large $|a|$
region could still be meaningful in e.g. a cosmological scenario.
This would be attributed to the fact
that ultimately higher energy scales must become important and hence may 
provide a stabilization of the theory, 
as has been speculated to be the case in the $N=1$ theories with a similar runaway solution at $T=0$
\cite{runawayn=1}. 

In the strong coupling region the stable points correspond to $|a_D| \propto g^2_DT$, so that the phase
of $a_D$ is irrelevant.
The position of the minima at $|a_D|$ is not likely to be modified by higher order terms in the 
low energy effective theory. However, the possible implications of the heavy particles, whose
mass is proportional to $|a|$, remain unclear. For instance, if the monopoles are strictly massless
the value of the
free energy is higher than at the point $M_m = (13g_D^2 T/12\pi )$. But since the slope at the
origin is zero there is no tendency to float
away from the massless case, and the theory seems to remain at
a local maximum. The ultimate fate of the massless theory must then be determined by the
effects of the heavy particles. 

Finally, we would like to comment on the different phases that exist at low temperatures.
If the temperature initially is much higher than any of the 
internal scales $\Lambda$ and $u$ in the microscopic
theory, we expect to have a nearly free gas of the ``electrical'' particles. As the temperature
is lowered, a description of the theory in terms of the low energy effective action
becomes valid. Whatever the nature is of
this transition, the theory will end up in one of the minima that exist at low temperatures.
The mechanism that drives the theory to a specific minimum presumably depends on how the scalar field
expectation value develops as the temperature is lowered. This is certainly a 
complicated process at an energy scale where standard methods such as perturbation theory is
not applicable. Without reliable
information it is obviously very dangerous to make any predictions, but it is not unreasonable
to expect that different regions in space could
fall into different minima. In that case, domain walls would form and remain until the temperature
vanishes. Needless to say, this scenario is presently by no means rigorous.   

\vskip 1cm
\noindent
{\bf Acknowledgments}
\renewcommand{\theequation}{6.\arabic{equation}}
\setcounter{equation}{0}
\vskip .5cm
\noindent
The author is grateful to T. H. Hansson for proposing the idea and
would like to thank T. H. Hansson, U. Lindstr\"{o}m and B. Sundborg
for interesting and useful 
discussions.

\vskip 1cm
\noindent
{\bf Appendix}
\renewcommand{\theequation}{A.\arabic{equation}}
\setcounter{equation}{0}
\vskip .5cm

In this appendix we fix our notation and illustrate the calculations of the free energy with
a simple example.

Starting in Minkowski space--time, our metric has the signature
\be
g^{\mu \nu} = {\rm diag} \left ( + \, -\, -\, - \right ) \ , 
\ee
and the antisymmetric tensor has $\epsilon_{0123} = +1$.
The $\gamma$--matrices satisfy $\{ \gamma^{\mu} , \, \gamma^{\nu} \} = 2g^{\mu \nu}$ and are given by
\be
\gamma^{\mu} = \left ( \matrix{0 \, & \sigma^{\mu} \cr \streck{\sigma}^{\mu} \, & 0} \right )
\hspace{1cm} \gamma_5 = \left ( \matrix{1\, & 0 \cr 0 \, & -1} \right ) \ ,
\ee
where $\sigma^{\mu} = (1 \, , \, \vec{\sigma})$, $\streck{\sigma}^{\mu} = (1 \, , \, -\vec{\sigma})$
and $\vec{\sigma}$ are the usual Pauli matrices. Also,
\be
\Sigma^{\mu \nu} = \frac{1}{4} \left [ \gamma^{\mu} , \, \gamma^{\nu} \right ] \ .
\ee

From the actions (\ref{effaction2})  and (\ref{dualcomponent})
we find the following propagators,
\be
&& \bullet \hspace{0.2cm} \mbox{fermions:} \hspace{0.5cm} \hspace{27pt} \frac{-i}{\pslash -m} \ ,
\nonumber \\
&& \bullet \hspace{0.2cm} \mbox{scalars:} \hspace{0.5cm} \hspace{36pt} \frac{i}{p^2-m^2} \ , \nonumber \\
&& \bullet \hspace{0.2cm} \mbox{vector bosons:} \hspace{0.5cm} \frac{-ig^{\mu \nu}}{p^2} \ ,
\ee
where $m$ is the possible particle mass.

To calculate the free energy, we first write the logarithm of the
partition function ${\rm {\cal Z}}$ in Minkowski space--time
and then make the following replacements,
\be
p_0 &\rightarrow& i\omega_n \nonumber \\
\int \frac{d p_0}{2\pi} &\rightarrow& iT\sum_{n=-\infty}^{\infty} \nonumber \\
2\pi \delta (p_0 ) &\rightarrow& -i\beta \delta_{\omega_n} \ , \label{replace}
\ee
where $\beta = T^{-1}$,
$\omega_n = 2n\pi T$ for bosons and  $\omega_n = (2n +1)\pi T$ for fermions. Labelling
the expression that follows when
the Minkowski rules and the replacements in equation (\ref{replace}) have been used by 
$(-i\beta V {\rm {\cal F}})$, with $V$ denoting the volume of the system, the
free energy density is,
\be
\Omega = -\frac{\log {\rm {\cal Z}}}{\beta V} = i {\rm {\cal F}} \ .
\ee

To give a simple illustration of the procedure outlined above, we will
calculate the contribution from the interaction term
\be
L_{{\rm eff}}^{(D)} = 
-\frac{g^2_D}{2} \left ( \phi^{\dagger} \phi \right ) \left ( \phi^{\dagger} \phi \right )
\ee
in the dual theory. We find,
\be
\Omega = g^2_D \left [ T\sum_n \int \frac{d^3 \! p}{(2\pi )^3} \, \frac{1}{(i\omega_n)^2 -
\vec{p}^{\, 2} - M_m^2 }\right ]^{2} \ .
\ee
We now write the sum over the Matsubara frequencies as a contour integral \cite{kapusta, lebellac},
\be
T\sum_nf(p_0 = i\omega_n) = \int_{-i\infty}^{i\infty}
 \frac{d p_0}{2\pi i} \, \left [ \frac{
f(p_0) +f(-p_0)}{2} \right ] \pm \int_{-i\infty+\epsilon}^{i\infty+\epsilon}
 \frac{d p_0}{2\pi i} \, \left [ \frac{
f(p_0) +f(-p_0)}{ e^{\beta p_0} \mp 1} \right ] , 
\ee
where the upper sign holds for bosons.
Omitting the temperature independent piece, we have
\be
\Omega = g^2_D \left [\int \frac{d^3 \! p}{(2\pi )^3} \, \frac{1}{\sqrt{\vec{p}^{\, 2}+M_m^2}}
\left ( \frac{1}{e^{\beta \sqrt{\vec{p}^{\, 2}+M_m^2}} -1} \right ) \right ]^2 \ , \label{ex}
\ee
by closing the contour in the right half--plane and picking up the pole at $p_0 = \sqrt{\vec{p}^{\, 2}+M_m^2}$.
When $T \gg M_m$ we can expand the remaining integral in $(M_m /T)$ by use of the following result,
\be
\int_0^{\infty} \frac{dx \, x^2}{\sqrt{x^2 +y^2}} 
\left ( \frac{1}{e^{\sqrt{x^2 +y^2}} -1} \right ) = \frac{\pi^2}{6} - \frac{y\pi}{2} +
\frac{y^2}{4} \left \{ \frac{1}{2} -\gamma - \log \left ( \frac{y}{4\pi} \right ) \right \} + 
{\cal {\rm O }} (y^3) \ .
\ee
Equation (\ref{ex}) then finally becomes,
\be
\Omega = \frac{g_D^2}{24} \left [ \frac{T^4}{6} - \frac{M_mT^3}{\pi} + \frac{M_m^2T^2}{2\pi^2}
\left \{ \frac{7}{2} -\gamma - \log \left ( \frac{M_m}{4\pi T} \right ) \right \} + {\cal {\rm O }} \! \! \left ( M_m^3 T 
\right ) \right ] \ .
\ee

\bibliographystyle{aip}
\bibliography{vacref}

\end{document}